# Single-sensor and real-time ultrasonic imaging using an AI-driven disordered metasurface

Wei Wang[1*], Jie Hu[2*], Jingjing Liu[1†], Yang Tan[1], Jing Yang[1], Bin Liang[1†], Johan Christensen[3†], Jianchun Cheng[1†]

[1]Key Laboratory of Modern Acoustics, MOE, Institute of Acoustics, Department of Physics, Collaborative Innovation Center of Advanced Microstructures, Nanjing University, Nanjing 210093, People's Republic of China

[2]Department of Information Science and Technology, Nanjing Forest University, Nanjing, 210037, China

[3]IMDEA Materials Institute, Calle Eric Kandel, 2, 28906, Getafe, Madrid, Spain

[*]These two authors contributed equally to this work.

[†]Correspondence and requests for materials should be addressed to Jingjing Liu (email: liujingjing@nju.edu.cn), Bin Liang (email: liangbin@nju.edu.cn), Johan Christensen (email: johan.christensen@imdea.es), or Jianchun Cheng (email: jccheng@nju.edu.cn)

**Abstract** - Non-destructive testing and medical diagnostic techniques using ultrasound has become indispensable in evaluating the state of materials or imaging the internal human body, respectively[1,2]. To conduct spatially resolved high-quality observations, conventionally, sophisticated phased arrays are used both at the emitting and receiving ends of the setup[3]. In comparison, single-sensor imaging techniques offer significant benefits including compact physical dimensions and reduced manufacturing expenses. However, recent advances such as compressive sensing[4] have shown that this improvement comes at the cost of additional time-consuming dynamic spatial scanning or multi-mode mask switching, which severely hinders the quest for real-time imaging. Consequently, real-time single-sensor imaging, at low cost and simple design, still represents a demanding and largely unresolved challenge till this day. Here, we bestow on ultrasonic metasurface with both disorder and artificial intelligence (AI). The former ensures strong dispersion and highly complex scattering to encode the spatial information into frequency spectra at an arbitrary location, while the latter is used to decode instantaneously the amplitude and spectral component of the sample under investigation. Thus, thanks to this symbiosis, we demonstrate that a single fixed sensor suffices to recognize complex ultrasonic objects through the random scattered field from an unpretentious metasurface, which enables real-time and low-cost imaging, easily extendable to 3D.

## Main

Owing to its high penetration depth and biocompatibility, imaging using ultrasound remains ubiquitous and pivotal in many areas such as medical ultrasonography and industrial crack and fatigue testing[5-7]. The chief idea to make structures or specimens visible through sound originates from military purposes for underwater sonar ranging and navigation, similar to echolocation of marine animals or bats. Firestone's supersonic reflectoscope and Dussik's hyperphonography, where the first techniques to use ultrasound as a means to detect internal flaws in metal castings and visualizing the ventricles of a human brain, respectively[8].

Ever since its infancy, ultrasonic imaging has come far. Today, phased arrays are typically used to probe complex geometries and a variety of material failures, as well as to scan inner organs for health diagnostics. Furthermore, specialized robotic surgeries employ real-time ultrasonic imaging for assisted localization and visualization[9,10]. Conventional ultrasonic imaging techniques usually utilize sophisticated phased arrays at both the emitting and receiving ends to quickly obtain the spatial information of objects[3,7,11-14], which due to unavoidable manufacturing costs and system size, impedes the integration and miniaturization of the imaging system. In comparison, existing single-sensor imaging techniques offer significant benefits including compact physical dimensions, systematical simplicity, and reduced manufacturing expense[4,15,16]. However, they usually require dynamic spatial scanning[16] or multi-mode mask switching[4,15] with time-consuming multiple measurements, making the real-time imaging cumbersome to attain. Although the number of required measurements can be decreased to a certain extent by using computational imaging algorithms such as compressive sensing, reconstructing an image still requires multiple measurements, and the ensuing effort of post-processing of the measured data increases significantly. Hence, the issue of time delay remains unresolved fundamentally. Beyond this, imaging harnessing compressive sensing also requires deterministic masks encoding, while at higher ultrasonic frequency, this demand becomes even more difficult to meet, both in terms of fabrication and precision. Therefore, real-time single-sensor imaging at low cost, whether in the realms of optics or acoustics, still stands as an immediate unresolved undertaking[17,18].

To overcome the above fundamental limitations, here we present a novel mechanism for real-time ultrasonic single-sensor imaging based on marrying an artificial structure[19-23] with artificial intelligence (AI)[24,25], which features fast imaging speed, low-cost fabrication, compact system size, and high imaging quality. Disordered media with waves scattering and interference patterns of uncontrollable complexity, are commonly considered detrimental to imaging. On the contrary, we aim at capitalizing on strong randomness, by extracting a wealth of the associated imaging information through AI. Our approach employs a single fixed transducer emitting a broadband ultrasonic signal, which passes through the object of interest whose scattered waves are further irregularly spread through the disordered metasurface in succession, before received by a single fixed receiver-sensor. Utilizing the rich complexity arising from the random parameter distribution and strong dispersion characteristics, the disordered metasurface can effectively encode the information of the objects' scattered field, by spreading throughout space the amplitudes of its spectral components detected by a single fixed sensor. By harnessing the remarkable learning and feature-extraction abilities of neural networks[26], the encoded signal captured by the sensor can be utilized to quickly reconstruct the geometry of the original object, without the need of prior knowledge of the metasurface's internal structure and material parameters. Hence, the metasurface used in our experiment, which was handcrafted without the aid of any sophisticated precision machining process at a cost much less than one dollar, proved sufficient to yield high imaging quality. It is believed that our imaging approach, based on low-cost disordered metasurfaces and AI has the potential to substitute highly sophisticated phased arrays for human or robotic ultrasonic inspection or medical visualization in the near future.

## Results

We conduct underwater ultrasonic experiments to emulate the working conditions in non-destructive testing and medical diagnostics. Our aim is to capitalize on the use of a disordered metasurface by exploiting its complex data encoding attributes, which we employ to recognize a variety of handwritten digits from the known Mixed National Institute of Standards and Technology (MNIST) database[27]. Before delineating this approach, we begin by discussing the general setup. As Fig. 1a shows, submerged in a water container, we have an ultrasonic transducer whose broadband signal undergoes scattering after insonifying an object of arbitrary shape. The distinctive scattered ultrasonic field strongly interacts with the disordered metasurface, whose randomness and dispersion characteristics enables effective signal encoding. Specifically, the randomness of the metasurface facilitates encoding in the spatial domain, while its dispersive response enables multiplexing the wave by its different spectral components. Consequently, the complex ultrasonic information of the scattered waves emanating an object, can be effectively captured, and interpreted via a single sensor at the metsurface's far-side, after which, we employ a neural network to extract the shape information of the object under study from the spectrum of the received signal. Through a series of matrix operations and nonlinear activation functions between layers, the neural network can efficiently construct an operational relationship between the input spectra and the output object shape, without necessitating any prior knowledge of the metasurface, requiring only a limited amount of data for training. Hence, in the absence of sophisticated material design requirements, media with a certain level of disorder, in combination with AI technology, can profoundly mitigate stringent system requisites and enable powerful imaging possibilities.

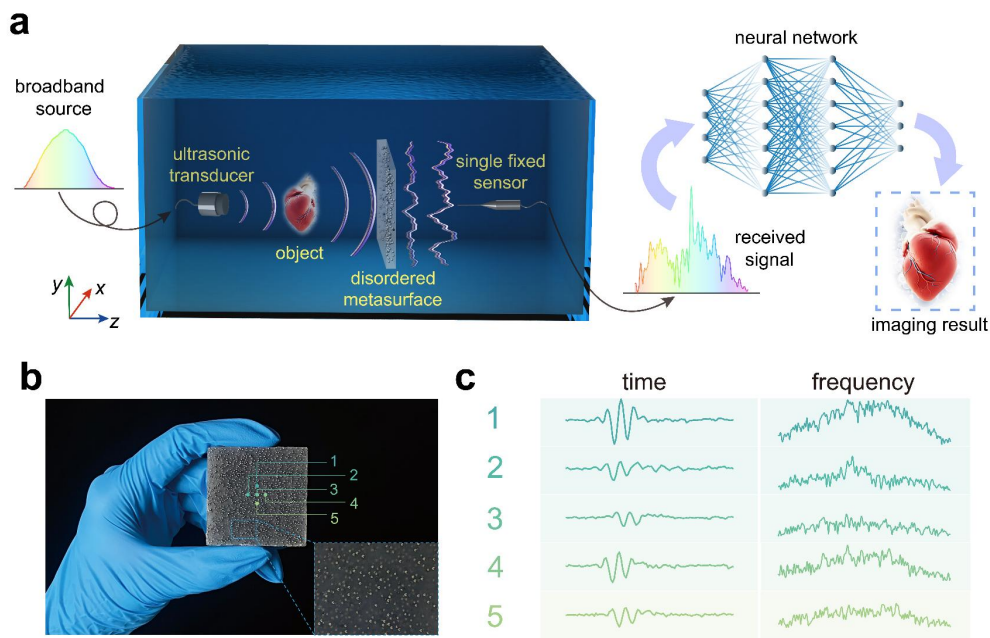

**Fig. 1| Single-sensor and real-time imaging system with AI-driven disordered metasurface. a**, Schematic of the proposed mechanism. At the emitting end, a fixed transducer generates broadband ultrasonic waves, which first undergoes scattering from the object and then impinges on the disordered metasurface. At the receiving end, a single fixed sensor captures the encoded signal, which serves as the input data of a neural network to reconstruct the image. **b**, Photograph of the fabricated disordered metasurface. **c**, Reference measured time-domain signals and the calculated frequency spectra at 5 particular points behind the metasurface (marked by 5 circle dots in **b**).

We begin by designing the ultrasonic disordered metasurface used in our imaging approach. The metasurface consists of an agar slab encapsulating randomly distributed steel beads, as shown in Fig. 1b, whose fabrication process is characterized by its remarkable simplicity (detailed fabrication process is discussed in Method).

Initially, an agar solution is subject to controlled heating until it reaches its boiling point, followed by a subsequent increase in temperature until a viscous state is attained. Next, a measured quantity of steel beads is blended into the solution, after which a cooling process is initiated, solidifying the material. Devoid of costly precision fabrication and advanced measurement instruments during this process, our metasurface solely requires facile random blending of steel beads. Thus, the metasurface costs much less than one dollar. To verify the randomness of the metasurface, we conducted an experiment where the metasurface is impinged by sinusoidal pulses with a center frequency of 2.5 MHz and a circle number of 2 (see Supplementary information for details). Measurements were taken at five points (green dots in Fig. 1b) near the back of the metasurface to analyze the transmitted signals. The temporal and frequency domain distributions of the receiver signals shown in Fig. 1c, clearly display strong disparity, which is an indicator of pronounced randomness.

To prepare the dataset for the neural network training, 700 handwritten digits from the MNIST Database are selected and engraved in metal plates as the objects to be imaged, with 70 samples for each digit from 0 to 9 (see Method and Supplementary information for details). We sequentially measured and recorded the signals received by a single fixed sensor placed behind the metasurface. Among the measurements, 600 samples were allocated for the training set, while 100 samples were designated for the test set, as shown in Fig. 2a. Figure 2b-2k present the normalized spectra of the scattered signals received by the sensor, which corresponds to the frequency-dependent fingerprints of the ten handwritten digits (0-9). Although these spectra appear highly intricate, seemingly hard to reveal any direct link and ability to extract the shape information of the digits, the spectra indeed exhibit certain variations but also some level of shape-similarity in relation to comparable handwritten digit, e.g., 4 and 9, as well as 5 and 6. This indicates that the disordered metasurface indeed is capable to scramble the scattered signals of objects, whose information is embedded in the spectral composition of the receiver data measured by a single sensor, which is the cornerstone of our imaging technique. This signal acts as the input dataset for training the neural network, while the labels for the neural network are the shapes of the objects. We employ a simple multilayer perceptron with two hidden layers, which are proved to be highly effective in imaging scattered objects such as handwritten digits (the detailed training process is discussed in Method).

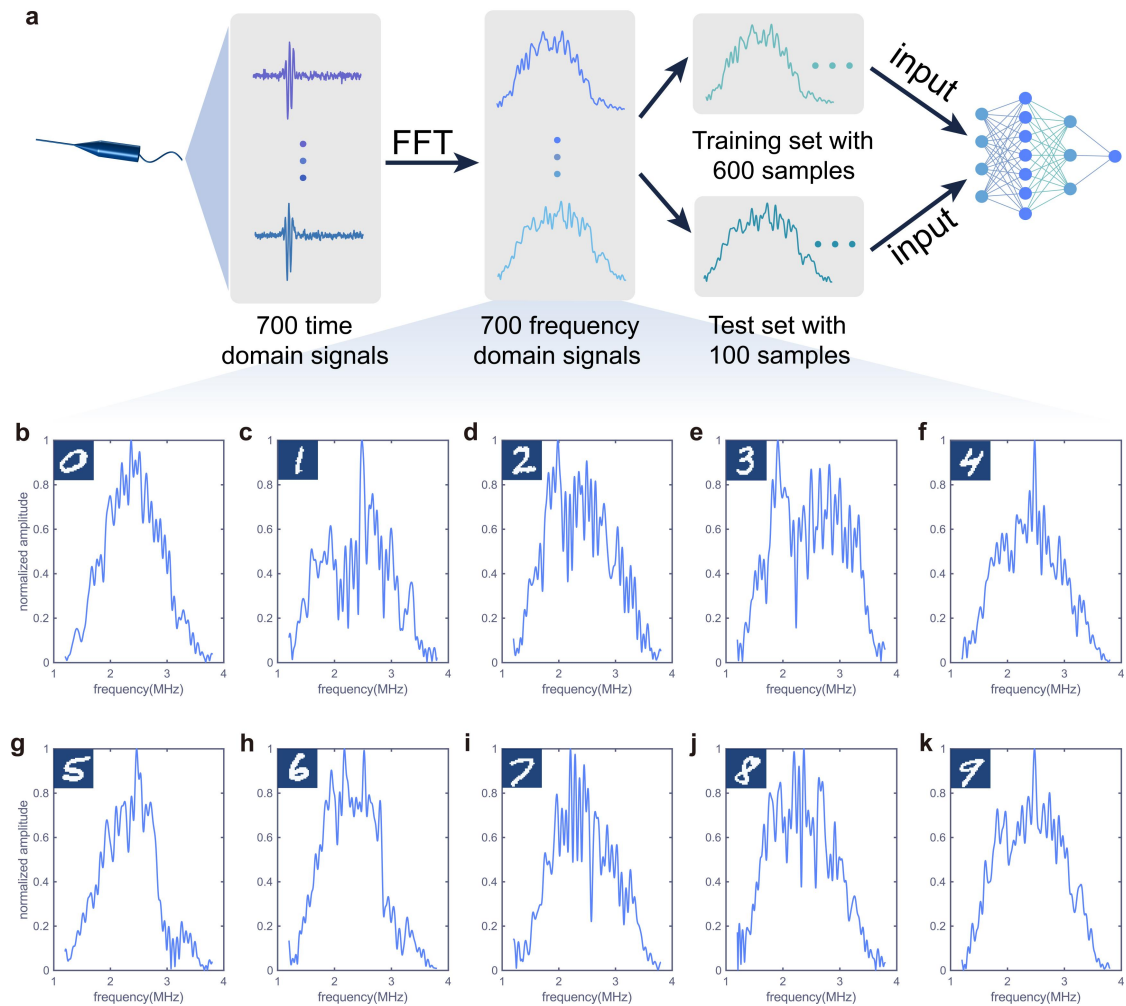

**Fig. 2| Dataset for the neural network training. a**, Process of the dataset preparation, for which 700 recorded signals comprise of different handwritten digits with 600 training sets and 100 test sets. The frequency spectrum of each signal is obtained through the fast Fourier transform and its normalization, which is the input data of the neural network. **b-k**, Typical normalized frequency spectra of different digits' scattering responses received by a single fixed sensor, for 0-9, respectively.

The detailed imaging results are shown in Figs. 3 and 4. Figure 3a depicts the flowchart of the whole imaging process. The emitted signals sequentially pass through the object and the disordered metasurface. The single fixed sensor receives the signals, which are then utilized to reconstruct the geometric shape of the object. Here we use the structural similarity (SSIM) between original objects and imaging results to evaluate the imaging quality. Through a simple neural network, the average SSIM for the training and test sets can reach 0.9870 and 0.7853, respectively. We select typical imaging results for 10 handwritten digits from the training and test sets, as shown in Fig. 3b-3d. Figures 3b and 3c compare the original shapes of the objects and the corresponding imaging results for the training set, respectively. The differences between these two sets of images are close to negligible, thereby implying a near-complete reconstruction of the shape information. Figures 3d and 3e compare the original shapes of the objects and the corresponding imaging results for the test set, respectively. Note, that these two sets of images also demonstrate a remarkable degree of resemblance, showcasing the high fidelity and accuracy of the imaging process. Therefore, it can be concluded that the neural network is capable of effectively decoding the encoded information based on the disordered metasurface scattering data and the comprehensive training provided by the training set. It is worth mentioning that once the neural network training has completed, the entire imaging process has completed within tens of microseconds, which can be made even faster by increasing the bandwidth of the emitted signal. This indicates the potential for real-time imaging of dynamic objects, such as a pulsating heart, which holds significant importance in practical medical applications.

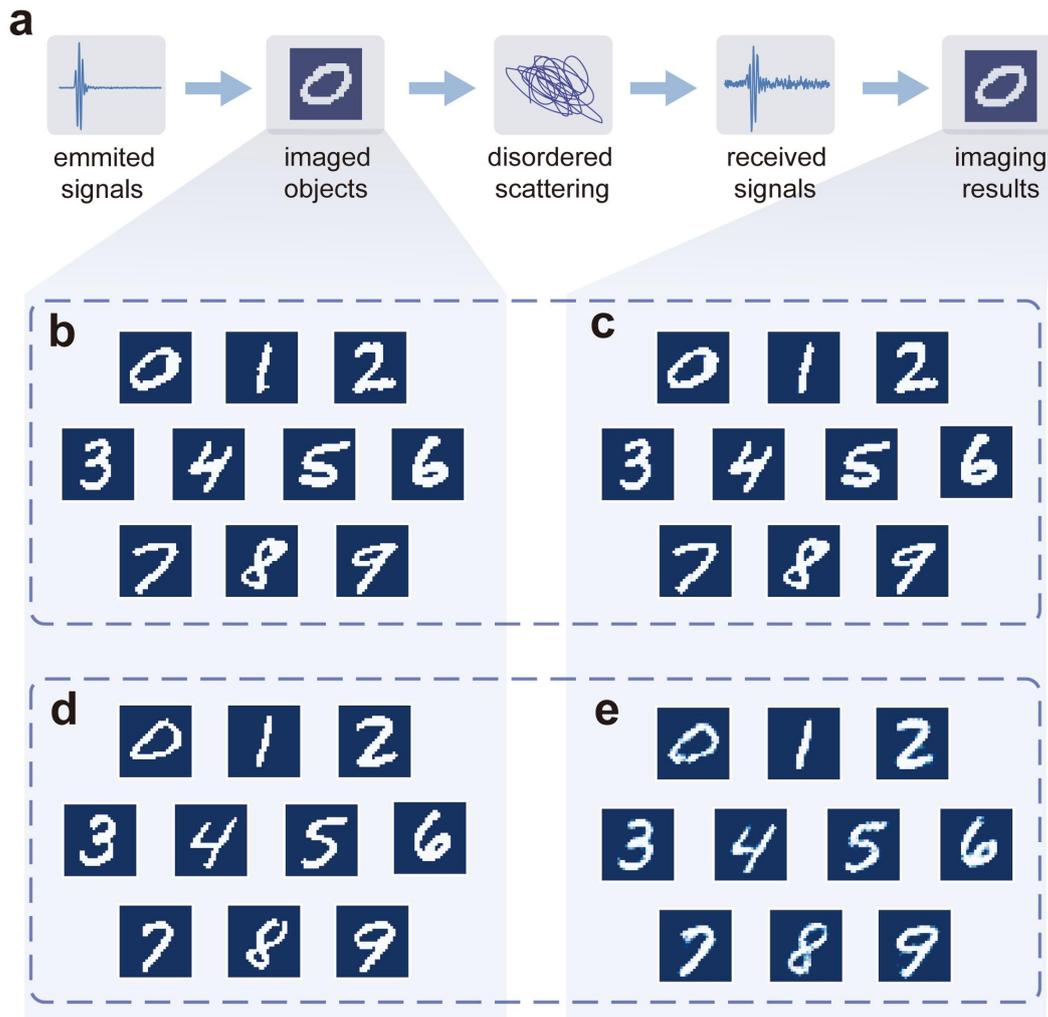

**Fig. 3| Typical imaging results for both the training and test dataset. a**, Flow chart of the imaging process. The emitted signals pass through the object and the disordered metasurface in sequence and are ensuringly captured to reconstruct the image of the original objects. **b**, Typical original objects of training set. **c**, Corresponding imaging results of objects in **b**. **d**, Typical original objects of test set. **e**, Corresponding imaging results of objects in **d**.

Figures 4a and 4b showcase the bubble charts that graphically depict the distribution of SSIM, quantifying the degree of similarity between the imaging results and their corresponding original objects, across each sample in both the training and test sets. In the training set, every sample demonstrates an SSIM exceeding 0.92, with a significant aggregation within the interval of 0.96 to 1.00. In the case of the test set, each sample's imaging results exhibit SSIM surpassing the threshold of 0.6, with no instance falling below this critical mark. It is worth noting, as elucidated in the

Supplementary information, that an SSIM value surpassing 0.6 signifies a commendable reconstruction of the shape information associated with the objects. This indicates the efficacy of our mechanism in achieving effective imaging for all samples, without any instances of poor imaging quality for individual samples. Such consistency and reliability in imaging performance hold crucial significance for practical applications, greatly increasing the diagnosis accuracy. Figure 4c illustrates the training process of the neural network, showing excellent convergence of the model. Although there exists overfitting issue, it does not affect the effectiveness of our mechanism and this problem tends to diminish as the sample size of the dataset increases. Furthermore, compared with common object recognition mechanisms based on neural networks[28], which can identify objects but may struggle to reconstruct the its precise shape, our method, without any compression of the information utilized for imaging, can reconstruct the geometry of the objects that have different shapes but belong to the same number, as shown in Fig. 4d.

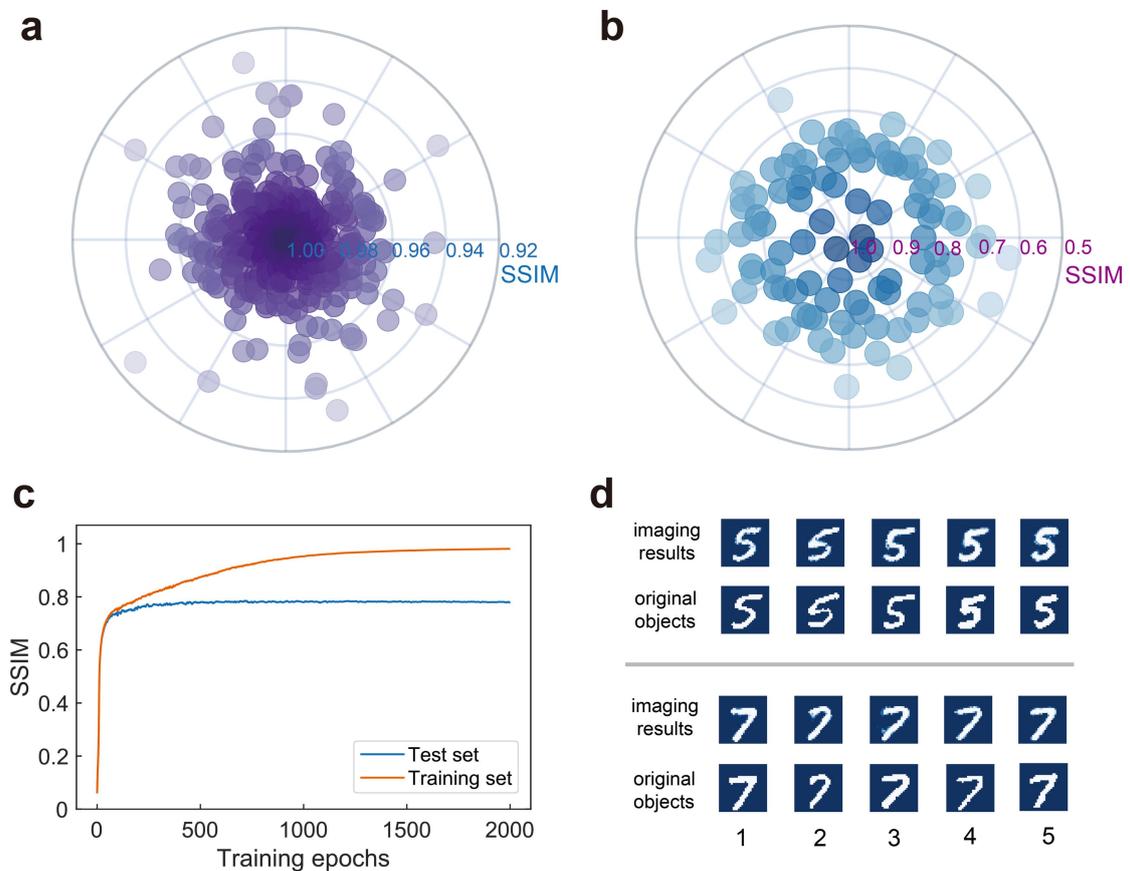

**Fig. 4| Overall imaging performance demonstration for our single-sensor**

**and real-time imaging system. a**, SSIM distribution of 600 digits in the training set. **b**, SSIM distribution of 100 digits in the test set. **c**, Variation curves of SSIM with the epoch of training. **d**, Two sets of imaging results of the same digits of different shapes in accompany with the original objects in the test set.

## Conclusion

In conclusion, we have demonstrated an approach that can realize real-time object imaging by a single fixed sensor with the help of an extremely low-cost disordered metasurface and AI technology. As a case example using this technique, we used an ultrasonic setting to experimentally detect and recognize different handwritten digits selected from the MNIST database. Our findings overcome a long-lasting bottleneck comprising a to date unresolved compromise between single-sensor and real-time imaging to ensure high image quality at low fabrication and machining costs. Combining a disordered metasurface and AI is universal and not limited to ultrasound. While we only demonstrated 2D handwritten digit imaging, it is entirely possible to extend it to complex objects in three-dimension (3D) as well and also considering other wave types. Relying only on wave randomness and dispersion, a plethora of artificially structured materials and compounds could be employed beyond the metasurface used in the present experiments. Thus, we anticipate that our approach will provide enough momentum to advance in real-time single-sensor applications for future low-cost devices interesting for industrial and medical applications and beyond.

**Acknowledgements**

The authors thank Dazhi Gao for fruitful discussions and for his contribution in the initial stage of the experiment. This work was supported by the National Key R&D Program of China (Grant Nos. 2022YFA1404402 and 2017YFA0303700), the National Natural Science Foundation of China (Grant Nos. 11634006 and 12174190), High-Performance Computing Center of Collaborative Innovation Center of Advanced Microstructures and a project funded by the Priority Academic Program Development of Jiangsu Higher Education Institutions. J.C. acknowledges support from the Spanish Ministry of Science and Innovation through a Consolidación Investigadora grant (CNS2022-135706).

## Data availability

The data that support the findings of this study are available from the corresponding authors on reasonable request.

## Code availability

The code used to calculate the results for this work is available from the corresponding authors on reasonable request.

## Author Contributions

Bin Liang and Wei Wang conceived the idea. Wei Wang, Jingjing Liu, Jie Hu, Bin Liang and Yang Tan prepared the samples and performed the experiment. Wei Wang, Jingjing Liu, Bin Liang, Johan Christensen contributed to the writing of the paper. Bin Liang, Johan Christensen, Jingjing Liu, and Jianchun Cheng supervised the entire study.

## Competing interests

The authors declare no competing interests.